\documentclass{aastex}
\usepackage{spr-astr-addons}
\usepackage{url}

\newcommand{\emaila}{wilhelm@mps.mpg.de}
\newcommand{\emailb}{ bnd.app@iitbhu.ac.in}

\def\deg{\hbox{$^{\rm o}$}}

\newcommand{\m}{ {\ \mathrm m} }
\newcommand{\s}{ {\ \mathrm s} }

\newcommand*\Del{\mathnormal{\Delta}}                 

\renewcommand{\vec}[1]{\mbox{\boldmath $#1$}}

\begin{document}

\title{Michelson--Morley experiment, Doppler effect, aberration of light and the aether concept}

\shorttitle{Doppler effect, aberration and the aether}
\shortauthors{K. Wilhelm and B.N. Dwivedi}

\author{Klaus Wilhelm}
\affil{Max-Planck-Institut f\"ur Son\-nen\-sy\-stem\-for\-schung
(MPS), Justus-von-Liebig-Weg 3, 37077 G\"ottingen, Germany \\ \emaila}

\and

\author{Bhola N. Dwivedi}
\affil{Department of Physics, Indian Institute of Technology
(Banaras Hindu University), Varanasi-221005, India \\ \emailb}

Last updated on \today

\vspace{1cm}

\begin{abstract}
After an overview of various citations relevant in the context of photon
propagation, the relativistic Doppler effect and the addition
theorem of velocities are first derived taking into account momentum and
energy conservation. Clocks and the aberration of
light are treated next, before the lengths of rods and the Lorentz
transformations are discussed.
The Michelson--Morley experiment is described at rest and in motion with
respect to a preferred aether system, first under the assumption of an
operation in vacuum. It is concluded that the aether concept is fully
consistent with the formal application of the Special Theory of
Relativity (STR). Whether a determination of the speed of the laboratory system
relative to the aether is possible, is considered next either for an operation
of the experiment in vacuum or in a medium with an index of refraction not
equal to one. In both cases, the answer appears to be negative.
\end{abstract}

\keywords{radiation: dynamics -- relativistic \\ processes -- techniques:
interferometric -- cosmology: theory}

\section{Introduction} 
\label{sec:intro}

The following statement \citep[][p.~126]{Ein17} highlights the importance
with regard to the Doppler effect\,--\,discovered by \citet{Dop42}\,--\,and the
aberration\,--\,first described by \citet{Bra27}:
\begin{enumerate} \item[ ]
{\small Whatever will eventually be the theory of electromagnetic processes,
the D\,o\,p\,p\,l\,e\,r principle and the aberration law will continue to be
valid, [...].}
\end{enumerate}
Since \citet{MicMor} carried out their famous experiment, the discussion
remains inconclusive on whether or not the vacuum is filled with some kind of
aether. A recent publication recounts this history \citep{KraOve}.

First we want to refer to early statements by Einstein and others concerning the
aether in the framework of the Special Theory of Relativity (STR)
\citep[][p.~413]{Ein08}:
\begin{enumerate} \item[ ]
{\small Only the concept of a light aether as carrier of the electric and
magnetic forces is not consistent with the theory discussed here; [...].}
\end{enumerate}
\citet{Lau08} discussed the Lorentz contraction in the context of the electron
theory \citep[cf.][]{Abr03} and the STR:
\begin{enumerate} \item[ ]
{\small Both theories [...] agree. The only difference concerns the shapes of
moving charges; one theory assumes that they are not affected, whereas the
other gives a contraction in the direction of motion. [...].}
\end{enumerate}
In this paper, we will assume that the shapes are not affected.
The alternative is the standard treatment based on the Lorentz contraction.
it would lead to slightly modified equations, but would not affect
the main results.

In response to critical remarks by \citet{Wie11}, \citet{Lau12} concluded
that the existence of the aether is not a physical, but a philosophical problem.
However, \citet[][p.~83]{Lau59} later differentiated between the
physical world and the mathematical formulation of STR:
\begin{enumerate} \item[ ]
{\small It owes its elegant mathematical guise Hermann MIN\-KOWS\-KI who [...]
introduced time as fourth coordinate on the same footing with the three
spatial coordinates to form a four-dimensional ``World''. However, this is only
a valuable mathematical trick; deeper insight, which some people want to see
behind it, is not involved.}
\end{enumerate}
\citet{Sch90} discussed Wiechert's support of the aether concept by presenting
unpublished material from about 1919 to 1922 containing
the following two statements:
\begin{enumerate} \item[ ]
{\small
Einstein's theory of relativity caused a sudden setback at the end of
1905 \citep[cf.][]{Ein05b}.
Admittedly Lorentz' Theory was formally very much
improved, and based on the options of the Lorentz transformations, Einstein
and others have erected both a beautiful and extended building that is without
doubt of great and lasting value for physics. However, in addition, an
epistemological foundation was added with new relativistic ideas leading to
inconsistencies with aether concepts.
\item[ ]
On the nature of the substratum of the world two ideas are in
conflict: the concept of spacetime und the aether.}
\end{enumerate}
In contrast to earlier statements, Einstein said at the end of his speech in
Leiden \citep{Ein20}:
\begin{enumerate} \item[ ]
{\small According to the General Theory of
Relativity (GTR) a space without aether cannot be conceived; [...].}
\end{enumerate}
For further quotations of the speech see \citet{Gra01} and \citet{Kos04}
for other statements by Einstein on the aether concept.

In 1927 Michelson confessed at a meeting in Pa\-sa\-de\-na in the
presence of H.A. Lorentz:
\begin{enumerate} \item[ ]
{\small `Talking in terms of the beloved old aether (which is now abandoned, though I
personally still cling a little to it), [...]' \citep[][p.\,342]{Mic28}.}
\end{enumerate}
\citet[][p.\,906]{Dir51} wrote in a letter to Nature:
\begin{enumerate} \item[ ]
{\small `If one examines the question in the light of present-day knowledge,
one finds that the {\ae}ther is no longer ruled out by relativity,
and good reasons can now be advanced for postulating an {\ae}ther.'}
\end{enumerate}
and
\citet{Bui58} stated in the summary:
\begin{enumerate} \item[ ]
{\small `There is therefore no alternative to the ether hypothesis.'}
\end{enumerate}
Considering these statements, it is only appropriate to revisit
the relationship of the mathematical formulation of the STR and its physical
contents as well as to reconsider the aether concept in this paper.
We will denote our laboratory system with~S.
It contains physical devices, such as rods, clocks,
photon\footnote{\citet{Ein05a} used the expressions ,,Energiequanten''
(energy quanta) and ,,Lichtquant'' (light quantum). The name ``photon'' was
later coined by \citet{Lew26}.} emitters and detectors. As far
as photons are concerned, we will frequently refer to the wave-particle
dualism \citep{Ein05a} by quoting their energy~$E_\nu = h\,\nu$, where
$h = 6.626\,070\,040 \times 10^{-34}$\,J\,s is Planck's constant (CODATA, 2014)
and, at the same time, characterize them by their frequency~$\nu$ and
wavelength~$\lambda$.
The System~S is either at rest in a putative aether system~S$_{\rm p}$
or moves with a velocity~$\vec{v}$ relative to S$_{\rm p}$.

The important questions are whether such a preferred aether system,
in which the propagation of photons is isotropic with a speed of light in
vacuum $c_0~=~299\,792\,458~\m\,\s^{-1}$~(exact) \citep[][p.\,22]{BIPM} is
compatible with physical experiments in laboratory systems and if\,--\,should
the answer be in the affirmative\,--\,experimental methods
can be devised to determine the speed~$v$.

Before we embark on this exercise, an interesting remark by
\citet[][pp.105/106]{Fer32} should be recalled\footnote{\label{Fermi}
In this citation:
 Energy levels are $w_{1,2}$ and $c$ is the speed of light in vacuum.}:
\begin{enumerate} \item[ ]
{\small `The change of frequency of the light emitted from a moving source is very
simply explained by the wave theory of light. But it finds also a simple,
though apparently very different, explanation in the light-quantum theory;
it can be shown that the Doppler effect may be deduced from the conservation
of energy and momentum in the emission process.\\
Let us consider an atom~A with two energy levels $w_1$ and $w_2$; the
frequency emitted by the atom when it is at rest is then
\begin{eqnarray}
\quad\quad\quad\quad\quad\quad\quad\quad\quad \nu = (w_2 - w_1)/h. \nonumber
\end{eqnarray}
Let us now suppose that the atom is excited and that it moves with
velocity~$V$; its total energy is then:
\begin{eqnarray}
\quad \quad \quad \quad  \quad \quad \quad \quad \quad
w_2 + \frac{1}{2}\,m\,V^2.         \nonumber
\end{eqnarray}
At a given instant the atom emits, on jumping down to the lower state, a
quantum of frequency~$\nu'$; the recoil of the emitted quantum produces a
slight change of the velocity, which after the emission becomes~$V'$; the
energy of the atom is then $w_1+\frac{1}{2}\,m\,V'^2$. We get therefore from
conservation of energy
\begin{eqnarray}
h\,\nu' = (w_2+\frac{1}{2}\,m\,V^2) - (w_1+\frac{1}{2}\,m\,V'^2)
=\quad\quad\quad \nonumber\\
\quad\quad\quad\quad\quad\quad\quad\quad h\,\nu + \frac{1}{2}\,m\,(V^2 - V'^2).
\nonumber
\end{eqnarray}
The conservation of momentum gives:
\begin{eqnarray}
\quad \quad \quad \quad \quad \quad \quad \quad \quad m\,\vec{V'} =
m\,\vec{V} - \frac{h\,\vec{\nu'}}{c}  \nonumber
\end{eqnarray}
where the bold face letters mean vectors. Taking the square we get:
\begin{eqnarray}
\quad \quad \quad \quad m^2V'^2 =
m^2V^2 + \frac{h^2\nu'^2}{c^2} - 2\,m\,V\,\frac{h\,\nu}{c}\,\cos\theta
\nonumber
\end{eqnarray}
$\theta$ being the angle between the velocity and the direction of emission.
From this equation and (76) we get, neglecting terms in $1/c^2$:
\begin{eqnarray}
\quad \quad \quad \quad \quad \quad \quad  \nu' =
\nu\,\left(1 + \frac{V}{c}\,\cos\theta\right)
\nonumber
\end{eqnarray}
which is the classic formula for the Doppler effect to a nonrelativistic
approximation.'}
\end{enumerate}
\section{Relativistic Doppler effect and addition theorem of
velocities}
\label{sec:Doppler}
Guided by Fermi's explanation of the Doppler effect,
we will now derive a relativistic formulation under the assumption of the
preferred system~S$_{\rm p}$.

An atom~A with mass~$m$ in its ground state at rest in S$_{\rm p}$ has an
energy of
%
\begin{eqnarray}
E_0 = m\,c^2_0 ~ .
\label{Eq:Rest_en}
\end{eqnarray}
The energy-momentum relation for a particle in motion is
\citep[cf.][]{Ein05a,Ein05c,Dir36,Oku89}:
%
\begin{eqnarray}
E^2 = m^2\,c^4_0 + \vec{p}^2c^2_0 = m^2\,c^4_0 + p^2c^2_0 ~ .
\label{Eq:Energy_S}
\end{eqnarray}
For the atom~A with a speed~$\vec{v}$ relative to System~S$_{\rm p}$, the
energy~$E$ in S$_{\rm p}$ can be found with the help of the momentum vector
%
\begin{eqnarray}
\vec{p} = \vec{v}\,\frac{E}{c^2_0} \quad \quad \quad
\left(~|\vec{p}| = p =\beta\,\frac{E}{c_0}~\right)
\label{Eq:momentum}
\end{eqnarray}
with $\beta = v/c_0$ and $|\beta| < 1$.
Combining Eqs.~(\ref{Eq:Energy_S}) and (\ref{Eq:momentum}) leads to
%
\begin{eqnarray}
E = \frac{m\,c^2_0}{\sqrt{1 - \beta^2}} = \gamma\,m\,c^2_0 ~ ,
\label{Eq:energy_gamma}
\end{eqnarray}
where
%
\begin{eqnarray}
\gamma = \frac{1}{\sqrt{1 - \beta^2}}
\label{Eq:Lorentz_factor}
\end{eqnarray}
is the Lorentz factor with $1 \le \gamma < \infty$ and
%
\begin{eqnarray}
E - E_0 = m\,c^2_0\,(\gamma - 1) = E_{\rm kin}
\label{Eq:kinetic}
\end{eqnarray}
is the kinetic energy.

If the atom is in an excited state with an excitation
energy~$\Del E = E_\nu= h\,\nu$, if  measured in the rest frame of the atom,
Eq.~(\ref{Eq:Energy_S}) reads
%
\begin{eqnarray}
(E^*)^2 = (m^*)^2c^4_0 + (p^*)^2c^2_0 ~,
\label{Eq:energy_nu_0}
\end{eqnarray}
with a mass \citep[cf.][p.\,641]{Ein05c}\citep[][p.\,394]{Lau20}
%
\begin{eqnarray}
m^* = m + \frac{\Del E}{c^2_0}
\label{Eq:m_h_nu}
\end{eqnarray}
and a momentum
%
\begin{eqnarray}
p^* = \beta\,\frac{E^*}{c_0} ~ ,
\label{Eq:Momentum}
\end{eqnarray}
cf. Eq.~(\ref{Eq:momentum}).
According to Eq.~(\ref{Eq:energy_gamma}), the energy can also be
expressed with the Lorentz factor~$\gamma$ as:
%
\begin{eqnarray}
E^* = \gamma\,m^*c^2_0 = \gamma\,(m\,c^2_0 + \Del E)~.
\label{Eq:en_gam_S}
\end{eqnarray}

During de-excitation of the atom, a photon will be emitted. For the sake of
simplicity, only directions parallel or anti-parallel to an $x$ axis will be
considered. Nevertheless two effects have to be evaluated: the motion of
System~S, in which atom~A is at rest, relative to the preferred
System~S$_{\rm p}$ and the recoil on the emitting atom.

Conservation of momentum {\em and}\,\footnote{Compare an important statement
by \citet[][pp.~127]{Ein17}:\\
If a light beam hits a molecule and leads to an absorption or emission
of the radiation energy~$h\,\nu$ by an elementary process, this will always
be accompanied by a momentum transfer of
$\frac{\displaystyle {h\,\nu}}{\displaystyle c}$ to the molecule, [...].
However one usually only
considers the e\,n\,e\,r\,g\,y exchange without taking the
m\,o\,m\,e\,n\,t\,u\,m exchange into account.}
energy in S$_{\rm p}$ requires
%
\begin{eqnarray}
p^\pm =
p^* \mp \frac{h\,\nu^\pm}{c_0}
\label{Eq:Momentum_abc}
\end{eqnarray}
and
%
\begin{eqnarray}
E^\pm = E^* - h\,\nu^\pm ~.
\label{Eq:E_star}
\end{eqnarray}
The notations $p^+$, $h\,\nu^+$ indicate the momentum and energy,
respectively, of a photon propagating in the positive $x$ direction and
$p^-$ and $h\,\nu^-$ the reverse.

The recoil can
conveniently be calculated by first assuming $v = 0$, i.e., the system~S
coincides with S$_{\rm p}$ and the photon emission is isotropic in both
systems.
With this assumption, Eqs.~(\ref{Eq:Momentum_abc}) and (\ref{Eq:E_star})
reduce to
%
\begin{eqnarray}
p^\pm_0 = \mp \frac{h\,\nu^\pm_0}{c_0}
\label{Eq:Momentum_red}
\end{eqnarray}
and
%
\begin{eqnarray}
E^\pm_0 = m\,c^2_0 + \Del E - h\,\nu^\pm_0 ~.
\label{Eq:E_star_red}
\end{eqnarray}
Applying Eq.~(\ref{Eq:Energy_S}) to this case gives:
%
\begin{eqnarray}
(E^\pm_0)^2 = m^2\,c^4_0 + (p^\pm_0)^2c^2_0 ~ .
\label{Eq:Energy_pm}
\end{eqnarray}
Eliminating $p^\pm_0$ and $E^\pm_0$ with the help of Eqs.~(\ref{Eq:Momentum_red})
and (\ref{Eq:E_star_red}), we get the the recoil redshift
after a short calculation and the (trivial) result that
it does not depend on the direction of the emission in this case:
%
\begin{eqnarray}
\nu^\pm_0 = \nu\,\frac{m\,c^2_0 + \Del E/2}{m\,c^2_0 + \Del E}
\approx \nu\,\left(1 - \frac{1}{2}\,\frac{\Del E}{m\,c^2_0}\right) ~ ,
\label{Eq:recoil}
\end{eqnarray}
where the approximation is valid for small $\Del E/(m\,c^2_0)$.

In the general case with a speed $v \ne 0$ of S with respect
to the preferred System~S$_{\rm p}$, the photon
energy $h\,\nu^\pm$ and the momentum $h\,\nu^\pm/c_0$
must now be evaluated in S$_{\rm p}$, in which the
propagation is assumed to occur.

Taking the square of Eq.~(\ref{Eq:E_star}) gives\,--\,together with
Eq.~(\ref{Eq:Energy_S}) and the consideration that after the emission
the mass of the atom is again $m$\,--\,the relation:
%
\begin{eqnarray}
m^2\,c^4_0 + (p^\pm)^2c^2_0 = \nonumber \\
(E^*)^2 - 2\,E^*h\,\nu^\pm + (h\,\nu^\pm)^2 ~.
\label{Eq:energy-pm}
\end{eqnarray}
The elimination of the momentum and energy terms using Eqs.~(\ref{Eq:m_h_nu})
to (\ref{Eq:Momentum_abc}) leads after a lengthy calculation\footnote{See
Appendix~A.} to
%
\begin{eqnarray}
\Del E~(m\,c^2_0 + \frac{\Del E}{2}) =
h\,\nu^\pm \,\gamma\,(m\,c^2_0 + \Del E)\,(1 \mp \beta)~,
\label{Eq:step_1}
\end{eqnarray}
and finally, with $\Del E= h\,\nu$ in the rest system of the emitter,
to the result that the relativistic Doppler shift depends on the direction
and the emission becomes anisotropic:
%
\begin{eqnarray}
\nu^\pm= \nu\,\frac{m\,c^2_0 + \Del E}{m\,c^2_0 +
\Del E}\,\frac{\sqrt{1 - \beta^2}}{1 \mp \beta} =
\nu^\pm_0\,\displaystyle{\sqrt{\frac{1 \pm \beta}{1 \mp \beta}}} ~ ,
\label{Eq:recoil_2}
\end{eqnarray}
where the definition of $\nu^\pm_0$ agrees with that in Eq.~(\ref{Eq:recoil}).
In what follows, we will generally neglect any recoil, for instance, by
employing the M\"o{\ss}bauer effect \citep{Moe58} to obtain a very large
effective mass in Eqs.~(\ref{Eq:recoil}) and (\ref{Eq:recoil_2}).
Eq.~(\ref{Eq:recoil_2}) then is equivalent to the relativistic Doppler
equation which \citet[][p.\,902]{Ein05b} derived for the separation
of a detector with constant speed~$v = \beta\,c_0$ relative to an emitter:
%
\begin{eqnarray}
\nu^- = \nu\,\sqrt{\frac{1 - \beta}{1 + \beta}} ~.
\label{Eq:doppler12}
\end{eqnarray}
The Doppler effect followed in Einstein's treatment
from the application of the Lorentz transformations
\citep[cf.][p.\,1505]{Poi05} to Lorentz' electrodynamics
\citep{Lor95,Lor03},
whereas Eq.~(\ref{Eq:recoil_2}) is a consequence of the momentum and energy
conservation.\footnote{Note that the energy difference between $h\,\nu$ and
$h\,\nu^\pm$ is compensated by a change of the kinetic energy of the
emitter, cf. \citet{Fer32} for $\gamma \approx 1$
and Eqs.\,(\ref{Eq:kinetic}) to (\ref{Eq:recoil_2}) for a relativistic
calculation, where Eq.~(\ref{Eq:en_gam_S}) shows that the term
$\gamma\,\Del E$ contributes to the energy of the moving excited atom
and is available during the emission process.}

The formulation of the detection of the photons with frequencies~$\nu^\pm$
would have required a similar treatment, but is simplified by assuming no
recoil and $\nu = \nu^\pm_0$.
If the detector is\,--\,together with the emitter\,--\,at rest in S
and, therefore, also moving in S$_{\rm p}$ with $v$,
the reverse of Eq.~(\ref{Eq:recoil_2}) shows that the
energy~$h\,\nu$ will be absorbed:
%
\begin{eqnarray}
\nu^\pm\,\displaystyle{\sqrt{\frac{1 \mp \beta}{1 \pm \beta}}} =
\nu\,\displaystyle{\sqrt{\frac{1 \pm \beta}{1 \mp \beta}}}
\displaystyle{\sqrt{\frac{1 \mp \beta}{1 \pm \beta}}} = \nu ~.
\label{Eq:inverse}
\end{eqnarray}

It is noteworthy that an iterative application of Eq.~(\ref{Eq:recoil_2})
(again with the simplification $\nu^\pm_0 = \nu$) will yield the velocity
addition theorem for parallel velocities\footnote{Einstein's original notation
of the velocity addition theorem for parallel velocities is:\\
%
\begin{eqnarray}
U = \frac{v + w}{1 + \displaystyle{\frac{v\,u}{V^2}}}  \nonumber
\end{eqnarray}
\citep[][p.\,905]{Ein05b}.
Einstein denotes the speed of light in vacuum by $V$ here and in the next but
one citation. $w$ represents a speed and not an energy
level, cf. Footnote~\ref{Fermi}. \citet{Mer84} argued that such a theorem can
be proven without involving light and that it would be consistent with an
aether at rest.}. For later applications, we write it in the form:
%
\begin{eqnarray}
w_y =
\frac{u_x + v}{1 + \displaystyle{\frac{u_x \,v}{c^2_0}}}~~~~~~{\rm or}~~~~~~\beta_y
= \frac{\beta_x + \beta}{1 + \beta_x\,\beta}~,
\label{Eq:Additiontheorem}
\end{eqnarray}
where $\beta_x = u_x/c_0$ and $\beta_y = w_y/c_0$ with indices $x$ and $y$ as
required for a unique formulation.

To prove the above statement, e.g., for positive velocities, we apply the Doppler
Eq.~(\ref{Eq:recoil_2}) first with $\beta_0 = u_0/c_0$ and then with
$\beta = v/c_0$, cf. Eq.~(\ref{Eq:momentum}):
%
\begin{eqnarray}
\nu^+ = \nu\,\displaystyle{\sqrt{\frac{1 + \beta_0}
{1 - \beta_0}}\,\sqrt{\frac{1 + \beta}{1 - \beta}}}
= \nu\,\sqrt{\displaystyle{\frac{1 + \displaystyle{\frac{\beta_0 + \beta}
{1 + \beta_0\,\beta}}}{1 - \displaystyle{\frac{\beta_0 + \beta}
{1 + \beta_0\,\beta}}}}} ~.
\label{Eq:cal}
\end{eqnarray}
Consequently, we obtain $\nu^+$ from
%
\begin{eqnarray}
\nu^+ = \nu\,\displaystyle{\sqrt{\frac{1 + \beta_3}{1 - \beta_3}}} ~,
\label{Eq:Add_plus}
\end{eqnarray}
where $\beta_3$ is
%
\begin{eqnarray}
\beta_3 = \frac{w_3}{c_0} = \frac{\beta_0 + \beta}{1 + \beta_0\,\beta}
\label{Eq:Additiontheorem_2}
\end{eqnarray}
consistent with Eq.~(\ref{Eq:Additiontheorem}).
The theorem thus also follows from energy and momentum conservation during the
photon emission. The derivation of Eq.~(\ref{Eq:cal}) assumed
that $|\beta_0| < 1$ and $|\beta| < 1$. If either $|\beta_0|$ or
$|\beta|$ is approaching the limit~1, $|\beta_3|$ also goes to 1.

\section{Clocks and aberration}
\label{sec:C_D_A}
In the previous section, we have postulated that the preferred System~S$_{\rm p}$
exists with an isotropic speed of light in vacuum of $c_0 = \nu\,\lambda$,
where $\nu$ and $\lambda$ are the frequency and wavelength of an
electromagnetic wave.

This assumption is consistent with the more general synchronization scheme
of many clocks at rest in an inertial system by \citet[][p. 415]{Ein08}
in order to define a time required by physical applications\footnote{In this
citation,
Einstein uses $c$ for the speed of light in vacuum.}:
\begin{enumerate} \item[ ]
{\small Let two points $A$ and $B$ with a separation~$r$, at rest in a
coordinate system, be equipped with clocks. If the clock at $A$
indicates~$t_A$, when a light beam propagating through the vacuum in the
direction~$A\,B$ reaches point~$A$, and if $t_B$ is the reading of clock~$B$,
when the beam arrives at~$B$, then it should always be $r/(t_B - t_A) = c$,
whatever might be the movements of the emitting source or other bodies.}
\end{enumerate}
What is a clock? The next two statements suggest that \citet{Ein07}
considered in most cases atomic oscillators as clocks:
\begin{enumerate} \item[ ]
{\small Mr. J. \citet{Sta06} demonstrated in a paper, which appeared last
year, that the moving positive ions of canal rays emit line spectra by
confirming and measuring the Doppler effect. He also performed investigations
with a view to find and study a second-order effect (proportional to
$(v/V)^2$). Since the experimental setup was not designed for this special
purpose, a definite result was not obtained.
\item[ ] In \citet[][p.~422]{Ein08} we find:\\
Since the oscillation process corresponding to a spectral line has
probably to be considered as an intra-atomic process, the frequency of which
is determined solely by the ion, we can regard such an ion as a clock
with a certain frequency~$\nu_0$.}
\end{enumerate}
However, it is obvious that a clock, in addition, needs a counter to number
the periods.
\begin{enumerate} \item[ ]
{\small `A clock therefore produces a time scale (its \emph{proper time}, in
relativistic terminology)' \citep{AudGui}}.
\end{enumerate}
\citet[][p.\,374]{IveSti41} were successful in measuring the second-order
effect mentioned by \citet{Ein07}, but summarized the observations by the
ambiguous statement:
\begin{enumerate} \item[ ]
{\small `The net result of this whole series of experiments is to establish
conclusively that the frequency of light emitted by
moving canal rays is altered by the factor $(1 - v^2/c^2)^{1/2}$.'}
\end{enumerate}
On page\,369 the authors refer to their earlier paper \citep{IveSti38},
where unfortunately conflicting equations
$\lambda = \lambda_0\,(1 - V^2/c^2)^{1/2}$ on page~216 and
$\nu = \nu_0\,(1 - V^2/c^2)^{1/2}$ on page~226 are given. The confusion
is augmented by the explanation of the second equation:
\begin{enumerate} \item[ ]
{\small `The present experiment establishes this rate as according to the
relation
$\nu = \nu_0\,(1 - V^2/c^2)^{1/2}$~,
where $\nu_0$ the frequency of the clock when stationary in the ether, $\nu$
its frequency in motion.'}
\end{enumerate}
In line with the results of Sect.~\ref{sec:Doppler}, the last phrase should
be modified: In an inertial system in which the atom is moving with $v$
an energy of
%
\begin{eqnarray}
h\,\nu = h\,\nu_0/\gamma
\label{Eq:excitation_energy}
\end{eqnarray}
will be emitted by the atom. The balance is taken up by the kinetic energy
of the atom.

\citet{Saaetal} confirmed the prediction of STR on a
level of $< 8 \times 10^{-8}$
by measuring the Doppler shifts of moving Li$^+$ ions in an Ives--Stilwell-type
experiment in line with the relativistic Doppler formulae:
%
\begin{eqnarray}
{\rm (a)}~~\nu_0 = \nu_{\rm r}^-\,\gamma\,(1 + \beta)~~{\rm and}~~
{\rm (b)}~~\nu_0 = \nu_{\rm r}^+\,\gamma\,(1 - \beta)~,
\label{Eq:Doppler_formulae}
\end{eqnarray}
where $\nu_0$ is the frequency in the frame~S of the ion and
$\nu_{\rm r}^-$ and $\nu_{\rm r}^-$ the frequencies in a frame~S$_{\rm p}$.
Adding the Eqs.\,(\ref{Eq:Doppler_formulae}a) and (b) shows that the
mean value of the energies $h\,\nu_{\rm r}^-$ and $h\,\nu_{\rm r}^+$ is
consistent with Eq.\,(\ref{Eq:excitation_energy}).

The general aberration relation is
%
\begin{eqnarray}
\cos \vartheta = \frac{\cos \vartheta^+ - \beta}{1 - \beta \cos \vartheta^+}~,
\label{eq:aberration}
\end{eqnarray}
where $\vartheta^+$ is the angle of light propagation in an inertial
System~S$_{\rm p}$
with respect to the direction of the motion of an inertial
system~S, and $\vartheta$ the corresponding angle in the moving system. It was
also obtained by \citet[][p.\,425]{Ein08} from the Lorentz
transformations. Resolving Eq.~(\ref{eq:aberration}) for $\cos\vartheta^+$
gives the reverse aberration formula:
%
\begin{eqnarray}
\cos \vartheta^+ = \frac{\cos \vartheta + \beta}{1 + \beta \cos \vartheta} ~ .
\label{eq:aberration_r}
\end{eqnarray}
For the special case of $\vartheta = 90\deg$, i.e.
$\cos \vartheta = 0$ it is $\cos \vartheta^+ = \beta$.
It can easily be
demonstrated that this follows from energy and momentum conservation as well.

Let an excited atom with a large mass~$m$ (so that its recoil can be
neglected) and an excitation energy~$h\,\nu$ move with a velocity~$\vec{v}$
in S$_{\rm p}$. Assume a photon emission perpendicular to $\vec{v}$ as seen
from the moving atom. Its energy is given by Eq.~(\ref{Eq:en_gam_S}) and its
momentum by Eq.~(\ref{Eq:Momentum}). The emitted photon has an energy of
$\Delta E = E^* - \gamma\,m\,c^2_2 = \gamma\,h\,\nu$ and, consequently,
the magnitude of its momentum vector is $\gamma\,h\,\nu/c_0$.
The momentum of the atom changes parallel to the velocity
by $\beta\,\gamma\,m\,c_0 - p^* = - \beta\,\gamma\,h\,\nu/c_0$.
Momentum conservation thus requires a momentum component of the photon parallel
to $\vec{v}$ of $\beta\,\gamma\,h\,\nu/c_0$. This yields together with its
magnitude
$\cos \vartheta^+ = (\beta\,\gamma\,h\,\nu/c_0)/(\gamma\,h\,\nu/c_0) = \beta$.
%

\begin{figure}
\centerline{\includegraphics[width=8.5cm,clip=1]{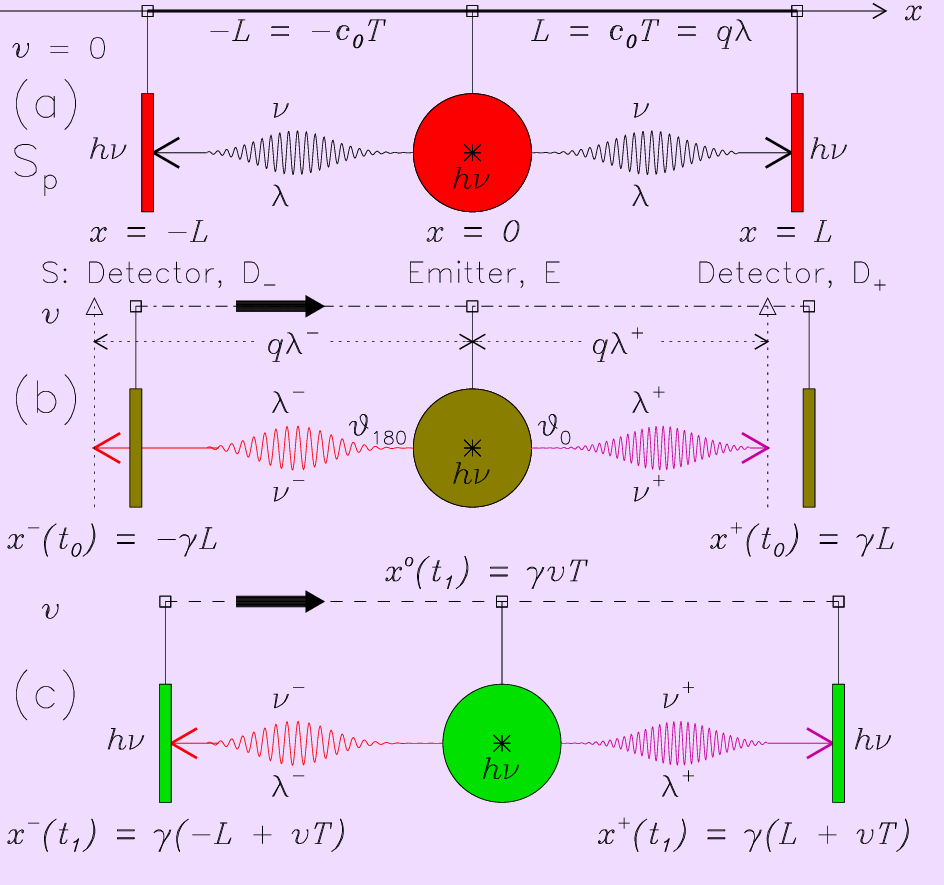}}
\caption{In Panel~(a), an emitter (circle) is located at $x_0 = 0$ and
radiates photons at time $t_0 = 0$ in $+$ or $-$ direction. They are
detected by the detectors~D$_-$ and D$_+$ (rectangles) at
$x = \pm L$, respectively, after travelling a
time~$T$ in the preferred system~S$_{\rm p}$\,--\,indicated by the violet
shading. The number of wavelengths in $L$ is denoted by $q$.
In Panels~(b) and (c), the arrangement of emitter and
detectors (System~S) is shown in S$_{\rm p}$, when
moving with a velocity~$v$ along the $x$-axis.
Panel~(b) illiustrates the situation at the emission of the photons;
together with their potential subsequent propagation in
S$_{\rm p}$ with $q$ wavelengths~$\lambda^\pm$ within
$T^\pm = T\,\sqrt{(1 \mp \beta)/(1 \pm \beta)}$.
Panel~(c) depicts the system at the time~$t_1$ of the detection of the photons,
when the emitter is located at $x^{\rm o}$.}

\label{Fig_Lorentz}
\end{figure}
%

%
\begin{figure}
\centerline{\includegraphics[width=8.5cm,clip=1]{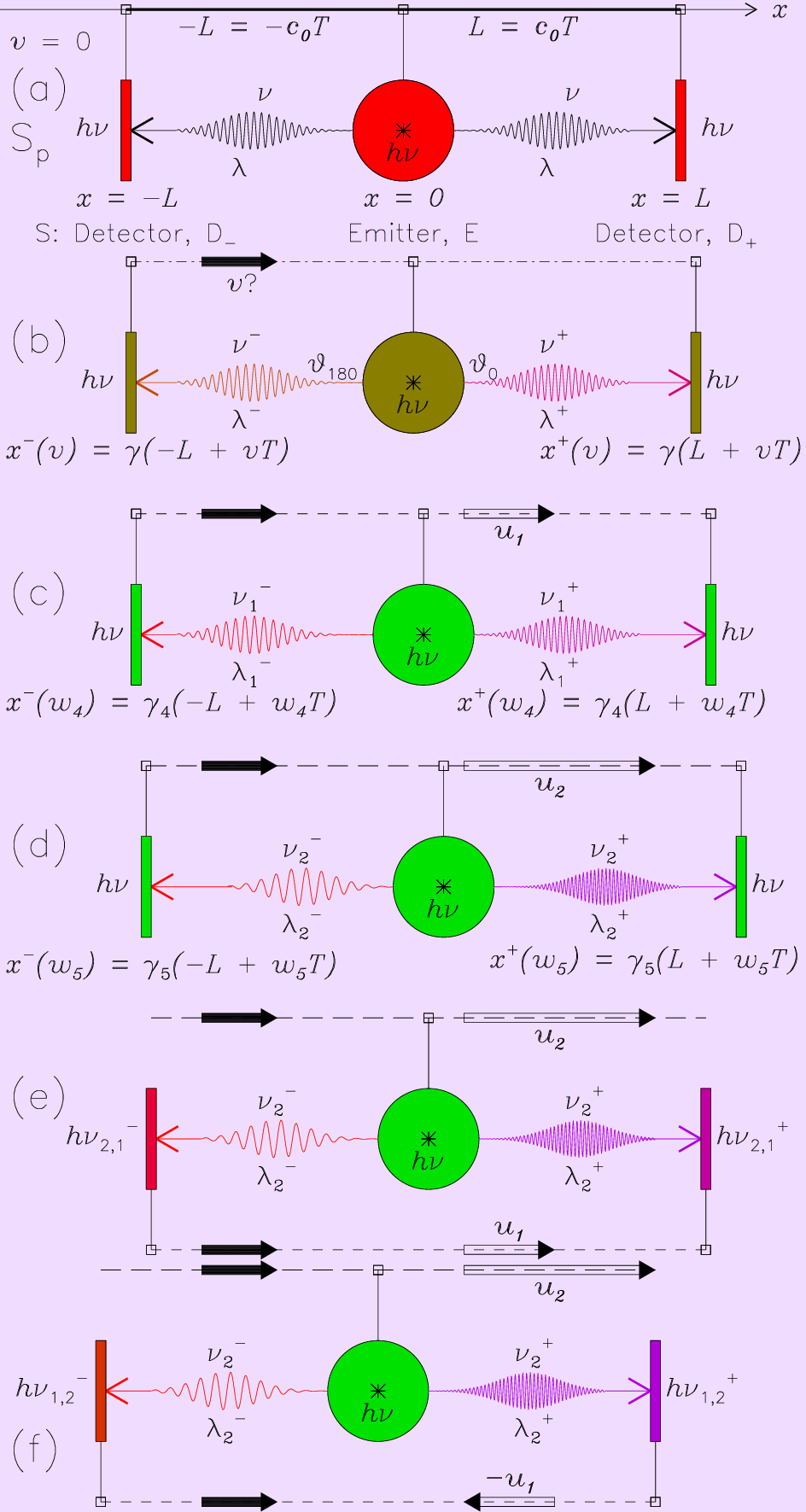}}
\caption{
As in Fig.~\ref{Fig_Lorentz}, a rod of length~$2\,L$ is equipped
with a source emitting photons with energy~$h\,\nu$ in the corresponding rest
frame of the emitter and two
detectors at the ends. In Panel~(a), the rod is at rest in
the preferred System~S$_{\rm p}$\,--\,indicated again by the violet
shading.
In Panel~(b), as in Fig.~\ref{Fig_Lorentz}(c), the emitter-detector frame~S
is now moving
with an unknown speed~$\vec{v}$ (solid arrow) relative to S$_{\rm p}$ in the
positive $x$ direction. Detection of the photons will then occur at $x^\pm(v)$
after times~$T^\pm(v)$ as seen from S$_{\rm p}$, in which the photon propagation
is assumed to happen. In Panels~(c) to (f), the emitter and the detectors are,
in addition, moving relative to~S with speeds of $\pm u_1$ and $u_2$,
respectively, which can directly be observed. Their speeds $w^\pm_1$ and
$w_2$ relative to S$_{\rm p}$ can then be calculated with the help of the
addition theorem, cf. Eq.~(\ref{Eq:Additiontheorem_2}). The detected energies
are calculated in the text.}
\label{Fig_Doppler}
\end{figure}
%

\section{Rods and the Lorentz transformations}
\label{sec:Rods}
\citet[][p.~392]{Edd23} felt that:
\begin{enumerate} \item[ ]
{\small `Size is determined by reference to material standards, and we must not
imagine that there can be any definition of size which dispenses with this
reference to material objects.'}
\end{enumerate}
Nevertheless two alternative methods have been used since: (1) The wavelength
of crypton~86 from 1960 to 1983 and (2) the present method based on clocks
and the speed of light in vacuum \citep[SI;][]{BIPM}.
Eddington had, however, qualified his conclusion by adding:
\begin{enumerate} \item[ ]
`No alternative method can
be accepted unless it has been proved to be equivalent to this.'
\end{enumerate}
Lorentz said at the Pasadena conference in 1927 about the contraction
hypothesis as an explanation of the Michelson--Morley experiment
\citep[][p.\,551]{Mic28}:
\begin{enumerate} \item[ ]
{\small `We are thus led to the ordinary theory of the experiment, which would make us
expect a displacement of the fringes, the absence of which is accounted for
by the well-known contraction hypothesis (Lorentz contraction).\\
Asked if I consider this contraction as a real one, I should answer ``yes.''
It is as real as anything that we can observe.'}
\end{enumerate}
A few years before this conference, \citet[][p.\,92]{Lau21} remarked on the
Lorentz contraction:
\begin{enumerate} \item[ ]
{\small If we set the body in motion without changing its shape, i.e. with
the old positions of the atoms, then is it possible that the forces could
vary in the same way as the forces between charges. [...] If they are of
electromagnetic nature, a rule derived by H.A. Lorentz says that the
Lorentz contraction results.}
\end{enumerate}

A rod of length~$2\,L$ aligned parallel to the $x$-axis of an inertial
system~S with an emitter of photons with an energy~$h\,\nu$ in S at the centre
and detectors at both ends is first assumed to be at rest in S$_{\rm p}$
in Fig.~\ref{Fig_Lorentz}(a). For photons propagating in
both directions along the rod, we can write with a certain number~$q$
%
\begin{eqnarray}
\pm L = \pm c_0\,T = \pm q\,\lambda = \pm q\,\frac{c_0}{\nu}~,
\label{Eq:L}
\end{eqnarray}
where $T$ is the travel time equal in both directions with $\pm c_0$ in the
preferred system. If this arrangement is now moving along the $x$-axis
with an (unknown) velocity~$\vec{v}$ relative to S$_{\rm p}$, the number~$q$ of
wavelengths can, at least in principle, be counted and, therefore, cannot
depend on the state of motion.
For this configuration, the longitudinal Doppler effect must be considered,
cf. Eq.~(\ref{Eq:recoil_2}). As shown in Sect.~\ref{sec:Doppler}, the
frequency shifts for the forward and backward directions follow directly from
the conservation of energy and momentum of the emitted photons.
We first discuss the forward direction:
%
\begin{eqnarray}
\nu^+ = \nu\,\gamma\,(1 + \beta) =
\nu\,\displaystyle{\sqrt{\frac{1 + \beta}{1 - \beta}}}
\label{Eq:nu_+}
\end{eqnarray}
with $c_0 = \nu^+\,\lambda^+$ in the preferred System~S$_{\rm p}$, we get
%
\begin{eqnarray}
\lambda^+  =
\frac{c_0}{\nu}\displaystyle{\sqrt{\frac{1 - \beta}
{1 + \beta}}}
\label{Eq:lambda_+}
\end{eqnarray}
and, with an invariant number\,$q$, a distance along this section of the
rod of
%
\begin{eqnarray}
L^+ = q\,\lambda^+ =
q\,\frac{c_0}{\nu}\,\displaystyle{\sqrt{\frac{1 - \beta}{1 + \beta}}} =
L\,\displaystyle{\sqrt{\frac{1 - \beta}{1 + \beta}}}
\label{Eq:L_+}
\end{eqnarray}
in the time~$T^+$, cf. Fig.\,\ref{Fig_Lorentz}(b).
Note that the photon travels after the emission in the preferred
System~S$_{\rm p}$ with speed~$c_0$, while the rod in the laboratory system~S
moves forward with~$v$. The question when and where the photon reaches the
front end of
the moving rod can be answered with the help of the paradox
``Achilles and the Tortoise'' formulated by Zeno of Elea. With
$\beta = v/c_0$ it will be at
%
\begin{eqnarray}
T_{\rm S} = T^+\,(1 + \beta + \beta^2 + \beta^3 + ...) =
\frac{T^+}{1 - \beta} = \gamma\,T ~
\label{Eq:T_S_plus}
\end{eqnarray}
and
%
\begin{eqnarray}
L_{\rm S} = L^+\,(1 + \beta + \beta^2 + \beta^3 + ...) =
\frac{L^+}{1 - \beta} = \gamma\,L ~.
\label{Eq:L_plus}
\end{eqnarray}

The propagation in the negative direction can be described by
%
\begin{eqnarray}
\nu^- = \nu\,\gamma\,(1 - \beta) =
\nu\,\displaystyle{\sqrt{\frac{1 - \beta}{1 + \beta}}} ~,
\label{Eq:nu_-}
\end{eqnarray}
and (with $c_0 = \nu^-\,\lambda^-$) a wavelength in the preferred
System~S$_{\rm p}$
 of
%
\begin{eqnarray}
\lambda^-  =
\frac{c_0}{\nu}\displaystyle{\sqrt{\frac{1 + \beta}
{1 - \beta}}} ~,
\label{Eq:lambda_-}
\end{eqnarray}
from which a propagation distance along this section of the rod of
%
\begin{eqnarray}
L^- = - q\,\lambda^- = - L\,\displaystyle{\sqrt{\frac{1 + \beta}{1 - \beta}}}
\label{Eq:L_-}
\end{eqnarray}
would follow. However, the photon now travels in the preferred
System~S$_{\rm p}$ with speed~$c_0$ in the negative $x$ direction,
while the laboratory system moves
forward with $+v$. The photon thus reaches the back end of the rod already at
%
\begin{eqnarray}
T^-\,(1 - \beta + \beta^2 - \beta^3 + ...) =
\frac{T^-}{1 + \beta} = \gamma\,T = T_{\rm S}
\label{Eq:T_S_minus}
\end{eqnarray}
and
%
\begin{eqnarray}
L^-\,(1 - \beta + \beta^2 - \beta^3 + ...) =
\frac{L^-}{1 + \beta} = - \gamma\,L = - L_{\rm S}~.
\label{Eq:L_minus}
\end{eqnarray}

When comparing the lengths $L^+$ and $L^-$ in Eqs.~(\ref{Eq:L_+})
and (\ref{Eq:L_-}), respectively, it is noteworthy that they differ in their
absolute values, but that the sum of the absolute values is
%
\begin{eqnarray}
L^+ + |L^-| =
L\left(\displaystyle{\sqrt{\frac{1 - \beta}{1 + \beta}}} +
\displaystyle{\sqrt{\frac{1 + \beta}{1 - \beta}}}\right) =  \nonumber \\
L\,[\gamma\,(1 + \beta) + \gamma\,(1 - \beta)] =
2\,\gamma\,L = 2\,L_{\rm S} ~,
\label{Eq:L_+_L_-}
\end{eqnarray}
i.e., exactly the length of $2\,L_{\rm S}$ resulting from
Eqs.~(\ref{Eq:L_plus}) and (\ref{Eq:L_minus}).
The photon emitter will be in the
middle of the rod. At time~$T_{\rm S} = \gamma\,T$,
the emitter will have moved to $v\,T_{\rm S} = v\,\gamma\,T$ and the detectors
to $\gamma\,(\pm L + v\,T)$. This situation is shown in
Fig.~\ref{Fig_Lorentz}(c).

We can now compare these findings with the results of a formal application
of the Lorentz transformations:\\
If the System~S is moving with speed~$v$ in the positive
$x$-direction of System~S$_{\rm p}$ and if both
systems agree at the time $t_0 = 0$, i.e. the events
$[x_0,t_0] = [0,0]$ and $[x^{\rm o},t^{\rm o}] = [0,0]$
coincide, then the inverse Lorentz transformations relate all other events
$[x,t]$ to $[x^\pm,t^\pm]$ by
%
\begin{eqnarray}
[x^\pm,t^\pm] =
[\gamma\,(x \pm v\,t),\gamma\,(t \pm \displaystyle{\frac{v\,x}{c^2_0}})]
\label{Eq:Lorentz}
\end{eqnarray}
\citep[cf.][]{Lor95,Lor03,Poi00,Ein05b,Jac01}.

For Systems~S and S$_{\rm p}$ the following space-time relations are obtained
under the assumption made in Sect.\ref{sec:intro} that the length~$L$ of the
rod in System~S does not change:
\begin{enumerate}
\item Emission of photon:\\
S: $[x_0,t_0] = [0,0]$ \\
S$_{\rm p}$$: [x^{\rm o},t^{\rm o}] = [0,0]$
\item Positions and times of detectors at photon emission:\\
S: $[x_{1,2},t_0] = [\pm L,0]$ \\
S$_{\rm p}$$: [x^\pm,t^\pm] = [\pm\gamma\,L,\pm \gamma\,\beta\,T]$
\item Positions and times of detectors at detection:\\
S: $[x_{1,2},t_1] = [\pm L,T]$ \\
S$_{\rm p}$$: [x^\pm,t^\pm] =
[\gamma\,(\pm L + v\,T),\gamma\,T\,(1 \pm \beta)]$
\item Positions and times of emitter at photon detection:\\
S: $[x_0,t_1] = [0,T]$ \\
S$_{\rm p}$$: [x^{\rm o},t^\pm] = [\gamma\,v\,T,\gamma\,T]$ ~.
\end{enumerate}
From Item~(iii) it follows
%
\begin{eqnarray}
\frac{\pm L}{T} = \frac{\gamma\,(\pm L + v\,T)}{\gamma\,T\,(1 \pm \beta)} =
\pm c_0 ~.
\label{Eq:Speed_light}
\end{eqnarray}
The conclusion can thus be drawn that Fig.~\ref{Fig_Lorentz} is in
agreement with the results obtained by applying in a formal way
the Lorentz transformations.

In Fig.~\ref{Fig_Doppler} some configurations are compiled to demonstrate
the relativistic longitudinal Doppler effect.
Inertial systems are assumed to
move with velocities of $\pm \vec{u}_1$ or
$\vec{u}_2$, respectively, relative to System~S (see open arrows in
Panels~(c) to (f)). This system is at rest in a putative aether
system~S$_{\rm p}$ in Panel~(a).
The solid arrows in Panels~(b) to (f)
indicate the unknown velocity~$\vec{v}$ with which
the systems are moving with respect to the aether, in addition to the
velocities relative to S.
In order to limit the complexity of the mathematical operations, it will be
assumed that all velocities are parallel or anti-parallel.
The total speeds~$w$ of the observational systems
in Panels~(c) to (f) expected relative to S$_{\rm p}$ can thus be obtained from
$|\vec{v}| = v$ and $\pm |\vec{u}_1| = \pm u_1$ or
$|\vec{u}_2| = u_2$ with the help of the velocity
addition theorem, cf. Eq.~(\ref{Eq:Additiontheorem_2}).
With $\beta^\pm_1 = u^\pm_1/c_0$ ($\beta^+_1 = \beta_1$; $\beta^-_1 = -\beta_1$)
and $\beta_2 = u_2/c_0$,
we get for Panels~(c)
and (d):
%
\begin{eqnarray}
\frac{w_4}{c_0} = \beta_4 =
\frac{\beta_1 + \beta}{1 + \beta_1\,\beta}
\label{Eq:beta_3}
\end{eqnarray}
and
%
\begin{eqnarray}
\frac{w_5}{c_0} = \beta_5 =
\frac{\beta_2 + \beta}{1 + \beta_2\,\beta} ~.
\label{Eq:beta_4}
\end{eqnarray}
The difference in Panel~(e) then is after evaluation using
Eqs.~(\ref{Eq:beta_3}) and (\ref{Eq:beta_4})
%
\begin{eqnarray}
\frac{\beta_5 - \beta_4}{1 - \beta_4\,\beta_3} =
\frac{\beta_2 - \beta_1}{1 - \beta_1\,\beta_2}
\label{Eq:beta_34}
\end{eqnarray}
and, therefore, is not dependent on $\beta$.
Similarly, we get in Panel~(f) with
%
\begin{eqnarray}
\frac{w^-_4}{c_0} = \beta^-_4 =
\frac{\beta^-_1 + \beta}{1 + \beta^-_1\,\beta} =
\frac{\beta - \beta_1}{1 - \beta_1\,\beta} ~,
\label{Eq:beta_3-}
\end{eqnarray}
and
%
\begin{eqnarray}
\frac{\beta_5 - \beta^-_4}{1 - \beta_5\,\beta^-_4} =
\frac{\beta_2 + \beta_1}{1 + \beta_1\,\beta_2} ~.
\label{Eq:beta_3-4}
\end{eqnarray}
The relativistic longitudinal Doppler shift thus is independent of $\beta$
in both cases.

In all Panels~(b) to (f), the frequencies of the propagating photons are
different from~$\nu$. Nevertheless, the detectors in Panels~(b) to (d)
measure the emitted frequency~$\nu$, because they are travelling with
the same speed as the emitter, cf. Eq.~(\ref{Eq:inverse}).
In Panels~(e) and (f), however, the relative motions of the detectors
relative to the emitter give
%
\begin{eqnarray}
\nu^\pm_{2,1} =
\nu\,\sqrt{\frac{1 \pm \beta_2}{1 \mp \beta_2}}\,
\sqrt{\frac{1 \mp \beta_1}{1 \pm \beta_1}}
\label{Eq:Det_2_1}
\end{eqnarray}
and
%
\begin{eqnarray}
\nu^\pm_{1,2} =
\nu\,\sqrt{\frac{1 \pm \beta_2}{1 \mp \beta_2}}\,
\sqrt{\frac{1 \pm \beta_1}{1 \mp \beta_1}} ~.
\label{Eq:Det_1_2}
\end{eqnarray}

A special application of Eq.~(\ref{Eq:Det_1_2}) might be instructive in showing
that the addition theorem can also be used to determine the total and kinetic
energies of a massive object relative to an observer. Let it move as emitter
in one direction with speed~$u_2$ relative to an inertial system, while the
observer and the detectors move in the opposite direction with~$-u_1$, cf.
Fig.~\ref{Fig_Doppler}(f). Note that this configuration is equivalent
to the iterative application of the Doppler effect with positive velocities
treated in Sect.~\ref{sec:Doppler}. Assume an electron-positron annihilation at the
emitter site with an energy release in its rest frame of
$E = 2\,m_{\rm e}\,c^2_0$. The energy absorbed by the detectors then is
%
\begin{eqnarray}
E^\pm =
m_{\rm e}\,c^2_0\,\sqrt{\frac{1 \pm \beta_2}{1 \mp \beta_2}}\,
\sqrt{\frac{1 \pm \beta_1}{1 \mp \beta_1}} \nonumber \\ =
m_{\rm e}\,c^2_0\,\sqrt{\frac{1 \pm \beta^1_2}{1 \mp \beta^1_2}} ~,
\label{Eq:Det_ann}
\end{eqnarray}
respectively, cf. Eqs.~(\ref{Eq:cal}) and (\ref{Eq:Add_plus}) with
%
\begin{eqnarray}
\beta^1_2 = \frac{\beta_1 + \beta_2}{1 + \beta_1\,\beta_2} ~.
\label{Eq:Additiontheorem_3}
\end{eqnarray}

The total absorbed energy in the detector frame is
%
\begin{eqnarray}
E^+ + E^- =  \nonumber \\
m_{\rm e}\,c^2_0\left(\displaystyle{\sqrt{\frac{1 + \beta^1_2}
{1 - \beta^1_2}}} +
\displaystyle{\sqrt{\frac{1 - \beta^1_2}{1 + \beta^1_2}}}\right) =
2\,\gamma^1_2\,m_{\rm e}\,c^2_0
\label{Eq:E_+_E_-}
\end{eqnarray}
with a Lorentz factor in the format
%
\begin{eqnarray}
\gamma^1_2 = \frac{1}{\sqrt{1 - (\beta^1_2)^2}} ~.
\end{eqnarray}
%
The kinetic energy in the observer system was
%
\begin{eqnarray}
E_{\rm kin} = 2\,m_{\rm e}\,c^2_0\,(\gamma^1_2 - 1).
\label{Eq: E_kin_gamma}
\end{eqnarray}

We have demonstrated with many examples that the application of momentum and
energy conservation during the emission and absorption of photons together
with the assumption of an aether as preferred System~S$_{\rm p}$ gives exactly
the same results obtained by formal application of the Lorentz transformations.
This answers the first question posed in Sect.~\ref{sec:intro} in the
affirmative. The second problem, however, to determine the speed of the
laboratory system relative to the aether could not yet be solved, because
all relations could be formulated without containing~$\beta = v/c_0$.

Even if we consider only the photon emission and
measure the recoil of the emitter as a function of~$v$ no effect will
be observed. Using
Eqs.~(\ref{Eq:Momentum}) to (\ref{Eq:E_star}), the problem can be
described in terms of energy and  momentum equations:
%
\begin{eqnarray}
\gamma\,(m\,c^2_0 + h\,\nu) = \gamma^\pm\,m\,c^2_0 + h\,\nu^\pm ~,
\label{Eq:en}
\end{eqnarray}
and
%
\begin{eqnarray}
\gamma\,\beta\,(m\,c^2_0 + h\,\nu) =
\gamma^\pm\,\beta^\pm\,m\,c^2_0 \pm h\,\nu^\pm ~.
\label{Eq:mo}
\end{eqnarray}
The photon terms can be eliminated by subtracting or adding the
Eqs.~(\ref{Eq:en}) and (\ref{Eq:mo}) written separately for $\beta^+$
and $\beta^-$. The results are
%
\begin{eqnarray}
\gamma\,(m\,c^2_0 + h\,\nu)\,(1 - \beta) =
\gamma^+\,m\,c^2_0\,(1 - \beta^+) \nonumber \\
\gamma\,(m\,c^2_0 + h\,\nu)\,(1 + \beta) =
\gamma^-\,m\,c^2_0\,(1 + \beta^-)
\label{Eq:sub_add}
\end{eqnarray}
and
%
\begin{eqnarray}
1 + \frac{h\,\nu}{m\,c^2_0} =
\displaystyle{\sqrt{\frac{1 + \beta}{1 - \beta}}\,
\sqrt{\frac{1 - \beta^+}{1 + \beta^+}}} =
\displaystyle{\sqrt{\frac{1 + \Del \beta^+}{1 - \Del \beta^+}}} \nonumber
\label{Eq:theorem}
\\
1 + \frac{h\,\nu}{m\,c^2_0} =
\displaystyle{\sqrt{\frac{1 - \beta}{1 + \beta}}\,
\sqrt{\frac{1 + \beta^-}{1 - \beta^-}}} =
\displaystyle{\sqrt{\frac{1 - \Del \beta^-}{1 + \Del \beta^-}}}
\label{Eq:Theorem}
\end{eqnarray}
with the substitutions
%
\begin{eqnarray}
\Del \beta^+ = \frac{\beta - \beta^+}{1 - \beta\,\beta^+} ~~{\rm and}~~
\Del \beta^- = \frac{\beta^- - \beta}{1 - \beta^-\,\beta}
\label{Eq:Delta}
\end{eqnarray}
in analogy to Eq.~(\ref{Eq:cal}).
Taking the square of Eq.~(\ref{Eq:theorem}) as well as of Eq.~(\ref{Eq:Theorem})
and setting
%
\begin{eqnarray}
\left(1 + \frac{h\,\nu}{m\,c^2_0}\right)^2 = B ~,
\label{Eq:B}
\end{eqnarray}
the equations can be solved for $\Del \beta^+$ and $\Del \beta^-$.
It can then, for instance,
be applied to the emission of Lyman~$\alpha$ by a hydrogen atom with the
assumptions of $v = 365$\,km\,s$^{-1}$\,--\,suggested by the asymmetry of
the cosmic background radiation \citep{Smo91}. We get
%
\begin{eqnarray}
c_0~\Del \beta^\pm =
\pm \,\frac{1 - B}{1 + B}\,c_0 = \mp 3.2573654~{\rm m}\,{\rm s}^{-1} ~.
\label{Eq:solve}
\end{eqnarray}
This is exactly the recoil speed of
$\mp 3.2573654~{\rm m}\,{\rm s}^{-1}$ for the Lyman~$\alpha$ emission
assuming $v = 0$\,km\,s$^{-1}$, because the equation is independent of $\beta$.


\section{Michelson--Morley experiment and the aether concept}
\label{sec:M_M_experiment}
%
\begin{figure*}
\centerline{\includegraphics[width=\textwidth,clip=1]{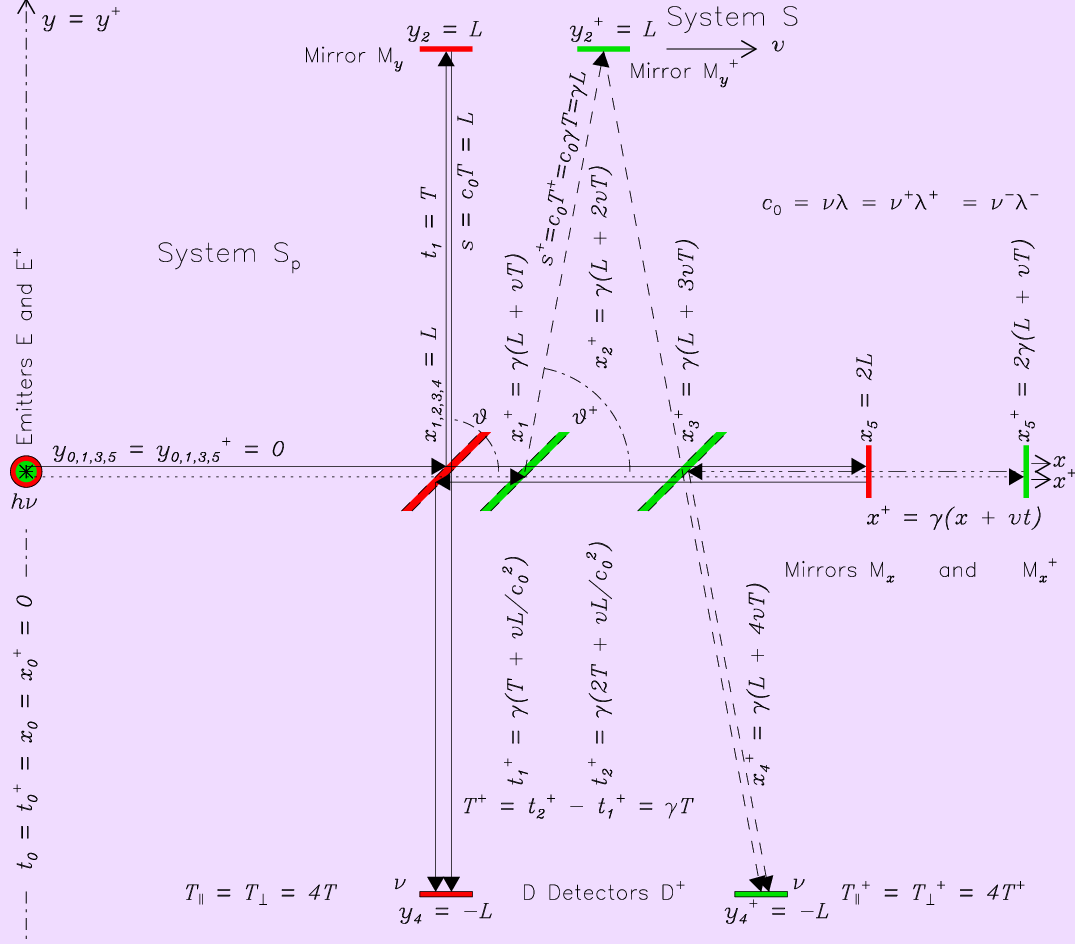}}
\caption{The Michelson--Morley experiment is outlined in the preferred
System~S$_{\rm p}$ by the red items emitter (radiating photons with an energy
of $h\,\nu$ in its rest frame), beam splitter, mirrors and detector, if the
apparatus in System~S is at rest in S$_{\rm p}$, as well as by green items
moving relative to System~S$_{\rm p}$ with a velocity~$\vec{v}$
parallel to the $x$-coordinate axes. The systems are assumed to coincide at
$t_0 = t^+_0 = 0$, when a photon in each of the systems is emitted. The
propagation of the photons occurs in System~S$_{\rm p}$. Their paths
are indicated by solid lines if $v = 0$ and their energy is~$E = h\,\nu$. If
$v \ne 0$, dotted lines signal an energy~$E^+ = h\,\nu^+$, dashed-dotted
lines an energy~$E^- = h\,\nu^-$, and dashed lines $E = \gamma\,h\,\nu$.
Further explanations are given in the text.}
\label{Fig_Mic_Mor}
\end{figure*}
%
\subsection{Operation in vacuum}
\label{Vacuum}

In Fig.~\ref{Fig_Mic_Mor}, the experimental setup in System~S is shown both at
rest in the preferred System~S$_{\rm p}$ with red optical elements and moving
with a speed~$v = \beta\,c_0$ in the positive $x$-direction with green optical
elements. The central beam splitter is shown twice at $[x^+_1,t^+_1]$ and at
$[x^+_3,t^+_3]$, when it is hit by the photon. Photons are radiated by an
emitter with energy~$E = h\,\nu$,
but if the emitter is moving with $v$, the photon
energy in S$_{\rm p}$ is according to Eq.~(\ref{Eq:recoil_2})
$E^+ = h\,\nu\,\sqrt{(1 + \beta)/(1 - \beta)}$ (dotted line). The receding mirror
at $x^+_5$ induces the inverse Doppler effect twice \citep[cf.][]{WilFro}
leading to
%
\begin{eqnarray}
E^- = h\,\nu\,\displaystyle{\sqrt{\frac{1 + \beta}{1 - \beta}}}\,
\frac{1 - \beta}{1 + \beta} =
h\,\nu\,\displaystyle{\sqrt{\frac{1 - \beta}{1 + \beta}}} ~,
\label{Eq:Dopp_Dopp}
\end{eqnarray}
shown by the dashed-dotted line.

The beam splitter at $x^+_1$, receding relative to the propagation direction
of the photons, causes an inverse Doppler effect.
The calculations at the end of Sect.~\ref{sec:C_D_A} can
be applied to this situation as follows: The photon is deflected in S by
90\deg, but travels with $E = \gamma\,h\,\nu$ to Mirror~M$^+_y$ and
Detector~D$^+$ (dashed lines) in S$_{\rm p}$ and,
consequently, the magnitude of its momentum vector is $\gamma\,h\,\nu/c_0$.
The momentum of the beam splitter changes parallel to the velocity
by $- \beta\,\gamma\,h\,\nu/c_0$.
Momentum conservation thus requires a momentum component of the photon parallel
to $\vec{v}$ of $\beta\,\gamma\,h\,\nu/c_0$. This yields together with its
magnitude\\
$\cos \vartheta^+ = (\beta\,\gamma\,h\,\nu/c_0)/(\gamma\,h\,\nu/c_0) = \beta$.

The beam splitter at $x^+_3$, advancing relative to the propagation direction
of the photon, will cause a corresponding Doppler effect and aberration
reflecting the photon also with $E = \gamma\,h\,\nu$ to Detector~D$^+$.
However, only $h\,\nu$ will be absorbed, because the detector momentum
change $\beta\,\gamma\,h\,\nu/c_0$ in $x$-direction has to be provided by the
photon in addition to the fractional energy. With $\cos \vartheta^+ = \beta$,
the photon with an energy of $\gamma\,h\,\nu/c_0$ just fulfills these
requirements.

Special mention should be made on the lengths of the inclined paths.
The $y$-component is $s = c_0\,T$ and the $x$-component
is $v\,(t^+_2 - t^+_1) = \beta\,\gamma\,c_0\,T$ thus
%
\begin{eqnarray}
s^+ = \sqrt{(c_0\,T)^2 + (\beta\,\gamma\,c_0\,T)^2} = \gamma\,c_0\,T
\label{Eq:Pyth}
\end{eqnarray}
is the geometric length in S$_{\rm p}$. A consequence is that from
$s = c_0\,T = q\,\lambda$ (cf. Fig.~\ref{Fig_Lorentz})
it follows that $s^+ = q\,\gamma\,\lambda$, confirming that the number
$q$ is not dependent on the motion. Eq.~(\ref{Eq:Pyth}) also shows that
the total length of the light path between the positions of the beam
splitter via the Mirror~M$^+_y$ is equal to the total length via the Mirror~M$^+_x$,
namely\\ $2\,\gamma\,L = (x^+_5 - x^+_1) + (x^+_5 - x^+_3)$.\\
Finally, it should be noted that
the times in both systems are related by $T^+ = \gamma\,T$ according to
Eq.~(\ref{Eq:Lorentz}).

All relations in Fig.~\ref{Fig_Mic_Mor} have been derived by momentum and
energy conservation of the emitted, propagating and absorbed photons and
optical elements. It is, however, important to note that they could have been
obtained in the framework of STR with the Lorentz transformations in
Eq.~(\ref{Eq:Lorentz}).
For the Systems~S and S$_{\rm p}$ in Fig.~\ref{Fig_Mic_Mor} the following
relations result:
\begin{enumerate}
\item Emission of photon:\\
$[x_0,t_0] = [0,0]$ \\ $[x^+_0,t^+_0] = [0,0]$
\item Position of beam splitter at first photon contact:\\
$[x_1,t_1] = [L,T]$ \\
$[x^+_1,t^+_1] = [\gamma\,(L + v\,T),\gamma\,(T + v\,L/c^2_0)]$
\item Position of mirror~M$^+_y$ at photon reflection:\\
$[x_2,t_2] = [L,2\,T]$ \\
$[x^+_2,t^+_2] = [\gamma\,(L + 2\,v\,T),\gamma\,(2\,T + v\,L/c^2_0)]$
\item Position of beam splitter at second photon contact:\\
$[x_3,t_3] = [L,3\,T]$ \\
$[x^+_3,t^+_3] = [\gamma\,(L + 3\,v\,T),\gamma\,(3\,T + v\,L/c^2_0)]$
\item Position of detector at photon detection:\\
$[x_4,t_4] = [L,4\,T]$ \\
$[x^+_4,t^+_4] = [\gamma\,(L + 4\,v\,T),\gamma\,(4\,T + v\,L/c^2_0)]$
\item Position of mirror~M$^+_x$ at photon reflection:\\
$[x_5,t_2] = [2\,L,2\,T]$ \\
$[x^+_5,t^+_5] = [2\,\gamma\,(L + v\,T),2\,\gamma\,(T + v\,L/c^2_0)]$
\end{enumerate}

\subsection{Operation in air}
\label{Air}

\citet{CahKit} have claimed that the interferometer experiment of
Michelson--Morley should give a null result only if operated under vacuum
conditions. \citet{BriHal} indeed obtained a null result in vacuum.
Historic experiments performed by \citet{MicMor} and \citet{Mil33}, however,
operated in air with an index of refraction at a wavelength of
$\lambda = 570$~nm of $n \approx 1.000\,2774$, which is only slightly depending on
the atmospheric pressure. They did, in most cases, not produce exact null
results.

In particular, \citet{Mil33} performed over decades many experiments  with
folded optical paths as long as 6406~cm, corresponding to a total light-path of
$112\,000\,000\,\lambda$. He found a maximum displacement of $0.152\,\lambda$
and converted it to an ``ether drift'' of 11.2~km s$^{-1}$
Can the observations accounted for by measurement uncertainties as is
generally done?

The propagation speed~$c$ of light in a transparent
body made out of a material with $n \ne 1$ moving with the speed~$v$ relative
to the observer was considered by \citet{Fre18} and \citet{Fiz51,Fiz60} with
the result that
%
\begin{eqnarray}
c = \frac{c_0}{n} \pm v \left(1 - \frac{1}{n^2}\right)~,
\label{Eq:Fresnel}
\end{eqnarray}
where $(1 - 1/n^2)$ is the so-called Fresnel aether drag coefficient,
indicating a partial drag of the aether by the moving body.
\citet{Fiz51}, however, was not convinced that these findings reflected the
actual physical process:
\begin{enumerate} \item[ ]
{\small `The success of the experiment seems to me to render the adoption of
Fresnel's hypothesis necessary, or at least the law which he found for
the expression of the alteration of the velocity of light by the effect
of motion of a body; for although that law being found true may be a
very strong proof in favour of the hypothesis of which it is only a
consequence, perhaps the conception of Fresnel may appear so
extraordinary, and in some respects so difficult, to admit, that
other proofs and a profound examination on the part of geometricians
will still be necessary before adopting it as
an expression of the real facts of the case.'
(cf. Comptes Rendus, Sept. 29, 1851)}
\end{enumerate}
More than 50~years later, \citet{Lau07} found that the light can be assumed to
be completely carried along by a body with $n \ne 1$, if the speeds
$c = c_0/n$ and $v$ are combined according to the addition theorem of
velocities. For parallel velocities $c$ and $v$, he obtained from
Eq.~(\ref{Eq:Additiontheorem}) a speed of
%
\begin{eqnarray}
w_\parallel^\pm = \frac{c \pm v}{1 \pm \displaystyle{\frac{v}{c_0\,n}}} =
c_0\,\frac{1 \pm n\,\beta}{n \pm \beta}
\approx \frac{c_0}{n} \pm v \left(1 - \frac{1}{n^2}\right) ~,
\label{Eq:Laue_para}
\end{eqnarray}
where the approximation, valid for small~$\beta$, is identical with
Fresnel's Eq.~(\ref{Eq:Fresnel}).
If $v$ is perpendicular to $c$ in System~S the theorem reads
%
\begin{eqnarray}
w_\perp = \displaystyle{\sqrt{c^2 + v^2\,\left(1 - \frac{1}{n^2}\right)}} =
\frac{c_0}{n}\,\sqrt{1 + \beta^2\,(n^2 - 1)} ~.
\label{Eq:Laue_perp}
\end{eqnarray}

Without taking the velocity addition theorem into account, \citet{CahKit}
deduced from Miller's and other observations drift speeds of about
400~km s$^{-1}$
We feel, however, that the velocity addition theorem has to be
applied in analysing the Michelson--Morley experiment. In
air, the time~$t_1$ is now $T = n\,L/c_0$ leading to the following relations:\\
The distance
%
\begin{eqnarray}
\Del x^+ = x^+_5 - x^+_1 = \gamma\,\left(L + v\,n\,\frac{L}{c_0}\right) =
\gamma\,L\,(1 + n\,\beta)
\label{Eq:x_+}
\end{eqnarray}
is traversed in the time
%
\begin{eqnarray}
\Del t^+ = \frac{\Del x^+}{w_\parallel^+} =
\gamma\,\frac{L}{c_0}\,(n + \beta) ~,
\label{Eq:t_+}
\end{eqnarray}
and in the reverse direction
%
\begin{eqnarray}
\Del x^- = x^+_5 - x^+_3 = \gamma\,\left(L - v\,n\,\frac{L}{c_0}\right) =
\gamma\,L\,(1 - n\,\beta)
\label{Eq:x_-}
\end{eqnarray}
in
%
\begin{eqnarray}
\Del t^- = \frac{\Del x^-}{w_\parallel^-} =
\gamma\,\frac{L}{c_0}\,(n - \beta) ~.
\label{Eq:t_-}
\end{eqnarray}
The corresponding relations via $[x^+_2,y^+_2]$ are:
The distance from $[x^+_1,y^+_1]$ to $[x^+_3,y^+_3]$  was
$\Del y^+ = 2\,\gamma\,L$ in vacuum, but might be different with $n \ne 1$.
So the task is to determine this distance. Several methods can be employed.
\begin{enumerate}
\item[a)]
Since the length~$L$ perpendicular to the velocity~$\vec v$ does not change,
we can together with the distance \\
$[x^+_1,x^+_2] = \gamma\,v\,n\,T = \gamma\,\beta\,n\,L$ calculate with the
help of Pythagoras' theorem the length of the light path between the
beam splitter and Mirror~$y^+_2$ in System~S$_{\rm p}$:
%
\begin{eqnarray}
s^+_n = L\,\sqrt{1 + \gamma^2\,\beta^2\,n^2} =
\gamma\,L\,\sqrt{1 + \beta^2\,(n^2 - 1)} ~.
\label{Eq:Pythagoras}
\end{eqnarray}
The distance to $[x^+_3,y^+_3]$ then is $2\,s^+_n$.
\item[b)]
With Eq.~(\ref{Eq:Laue_perp}) and $v$ we can find
%
\begin{eqnarray}
\cos \vartheta^+_n = \frac{n}{c_0}\,\frac{v}{\sqrt{1 + \beta^2\,(n^2 - 1)}} ~,
\label{Eq:speeds}
\end{eqnarray}
where $\vartheta^+_n$ is the angle $\vartheta^+$ in Fig.~\ref{Fig_Mic_Mor}
for an operation in air. Again we get:
%
\begin{eqnarray}
s^+_n = \frac{\gamma\,\beta\,n\,L}{\cos \vartheta^+_n} =
\gamma\,L\,\sqrt{1 + \beta^2\,(n^2 - 1)} ~.
\label{Eq:cos}
\end{eqnarray}
\item[c)]
\citet{Sheetal} have recently shown that their experiment supports Abraham's
concept of a momentum of light in media with $n \ne 1$ proportional to $1/n$.
Since in our case the emitter, the beam splitter and the air are moving
together with speed~$v$ relative to S$_{\rm p}$, a photon will be radiated
perpendicular to $\vec{v}$ with energy~$h\,\nu$ and
momentum~$h\,\nu/(n\,c_0)$ under the assumption of no recoil.
From Eqs.~(\ref{Eq:Momentum}) and (\ref{Eq:en_gam_S}) and the conservation of
energy and the $x$-component of the momentum, we get
%
\begin{eqnarray}
\tan\vartheta^+_n =
\frac{\displaystyle{\frac{h\,\nu}{n\,c_0}}}{\displaystyle{
\frac{\gamma\,h\,\nu\,\beta}{c_0}}} =
\frac{1}{\gamma\,n\,\beta}
\label{Eq:tan}
\end{eqnarray}
and with $\tan\vartheta^+_n = \sin\vartheta^+_n/\cos\vartheta^+_n$
%
\begin{eqnarray}
\cos \vartheta^+_n = \frac{n}{c_0}\,\frac{v}{\sqrt{1 + \beta^2\,(n^2 - 1)}} ~,
\end{eqnarray}
the same result as in b).
\end{enumerate}
Invoking all the three methods, we thus obtain
%
\begin{eqnarray}
2\,s^+_n = 2\,\gamma\,L\,\sqrt{1 + \beta^2\,(n^2 - 1)}
\end{eqnarray}
and
%
\begin{eqnarray}
\Del t_\perp = \frac{2\,s^+_n}{w_\perp } =
\frac{2\,\gamma\,L\,\sqrt{1 + \beta^2\,(n^2 - 1)}}
{\displaystyle{\frac{c_0}{n}\,\sqrt{1 + \beta^2\,(n^2 - 1)}}} =
\frac{2\,n\,\gamma\,L}{c_0} ~.
\label{Eq:t_perp}
\end{eqnarray}
We can now calculate for the different arms of the
interferometer
%
\begin{eqnarray}
\Del t^+ + \Del t^- - \Del t_\perp = 0
\label{Eq:t_diff}
\end{eqnarray}
and find that there is no delay. It therefore appears as if Cahill's and
Kitto's claim is not supported assuming von Laue's drag velocities.

\section{Discussion and conclusion}
\label{sec:concl}
An aether concept\,--\,required by GTR \citep{Ein24}\,--\,is not inconsistent
with STR, and allows us to interpret
the photon processes on the basis of momentum and energy conservation.
A determination of
the speed of a laboratory system relative to the aether does, however, not
seem to be possible, i.e., neither with operation in vacuum nor in air.

\section*{Acknowledgements}
This research has made extensive use of the Smithsonian Astrophysical
Observatory (SAO)/National Aeronautics and Space Administration (NASA)
Astrophysics Data System (ADS). Administrative support has
been provided by the Max-Planck-Institute for Solar System Research and the
Indian Institute of Technology (Banaras Hindu
University).

\appendix

\section{Derivation of the relativistic Doppler effect}
The evaluation of Eq.~(\ref{Eq:energy-pm}) using Eqs.~(\ref{Eq:m_h_nu}) to
(\ref{Eq:Momentum_abc}) can be achieved as follows:
%
\begin{eqnarray}
(E^*)^2 - 2\,E^*h\,\nu^\pm + (h\,\nu^\pm)^2 =
m^2\,c^4_0 + (p^\pm)^2c^2_0 = \nonumber \\
m^2\,c^4_0 + (p^*)^2c^2_0 \mp 2\,p^*\,c_0\,h\,\nu^\pm
+ (h\,\nu^\pm)^2 =
m^2\,c^4_0 + (E^*)^2\,\beta^2 \mp 2\,\beta\,E^*\,h\,\nu\pm
+ (h\,\nu^\pm)^2~.
\end{eqnarray}
Deletion of $(h\,\nu^\pm)^2$ from the first and last
lines gives:
%
\begin{eqnarray}
(E^*)^2 - 2\,E^*h\,\nu^\pm = \nonumber
m^2\,c^4_0 +(E^*)^2\,\beta^2 \mp 2\,\beta\,E^*\,h\,\nu^\pm ~. \nonumber
\end{eqnarray}
and
%
\begin{eqnarray}
(E^*)^2\,(1 - \beta^2) - 2\,E^*\,h\,\nu^\pm\,(1 \mp \beta) = m^2\,c^4_0~.
\end{eqnarray}
Substituting finally $E^*$ results in
%
\begin{eqnarray}
\Del E\,(m\,c^2_0 + \Del E/2) =
\gamma\,(m\,c^2_0 + \Del E)\,h\,\nu^\pm\,(1 \mp \beta)
\end{eqnarray}
after deletion of $m^2\,c^4_0$ on both
sides and noting that\\ $\gamma^2 = 1/(1 - \beta^2)$.
With $\Del E =h\,\nu$ in the emitter system the relativistic Doppler
formula follows.


\begin{thebibliography}{00}
\bibitem[\protect\citeauthoryear{Fizeau}{1860}]{Fiz60}
Fizeau, H., 1860,
On the effect of the motion of a body upon the velocity with which it is
traversed by light,
Phil. Mag., Series 4, 19, 245
\end{thebibliography}
\end{document}